\documentclass[twocolumn]{IEEEtran}  %%, draft
\usepackage{amssymb,amsmath}
\ifx\pdftexversion\undefined
  \usepackage[dvips]{graphics}
\else
  \usepackage[pdftex]{graphics}
\fi

\renewcommand {\langle}{(}
\renewcommand {\rangle}{)}

\newcommand {\mm} [4] {\arraycolsep=3pt
\left(\begin{array}{cc} #1 & #2 \\ #3 & #4
\end{array} \right)}
\newcommand {\mmm} [4] {\arraycolsep=3pt
\begin{array}{cc} #1 & #2 \\ #3 & #4
\end{array}}
\newcommand {\mmmm} [4] {
\left(\begin{array}{c|c} #1 & #2\\[0.8ex]\hline #3 & #4
\end{array} \right)}

\newcommand {\proj}{\mbox{proj}}
\newcommand {\note}{\noindent\hspace{1em}{\it Note: }}

\newtheorem{definition}{Definition}
\newtheorem{theorem}{Theorem}
\newtheorem{lemma}{Lemma}
\newtheorem{corollary}{Corollary}[lemma]

\title{Conditionally independent random variables}

\author{Konstantin Makarychev and Yury Makarychev
\thanks{Princeton University}
\thanks{E-mail: \{kmakaryc,ymakaryc\}@princeton.edu}
\thanks{This work was done while the authors were at Moscow State University.}
\thanks{Supported by Russian Foundation for Basic Research grant 01-01-01028.}
\thanks{
This work has been submitted to the IEEE for possible
publication. Copyright may be transferred without notice,
after which this version may no longer be accessible.}}

\begin{document}
\maketitle

\begin{abstract}
In this paper we investigate the notion of conditional
independence and prove several information inequalities for
conditionally independent random variables.
\end{abstract}


\section{Introduction}
Ahlswede, G\'acs, K\"orner, Witsenhausen and
Wyner~\cite{ahlswedekorner,ahlswedekornerinternet,gacskorner,witsenhausen,wyner}
studied the problem of extraction of ``common information'' from
a pair of random variables. The simplest form of this problem is
the following: Fix some distribution for a pair of random
variables $\alpha$ and $\beta$. Consider $n$ independent pairs
$\langle \alpha_1,\beta_1 \rangle, \dots, \langle \alpha_n,
\beta_n\rangle$; each has the same distribution as $\langle
\alpha,\beta\rangle$. We want to extract ``common information'' from
the sequences $\alpha_1,\dots\alpha_n$ and $\beta_1,\dots,\beta_n$,
\textit{i.e.}, to find a random variable $\gamma$ such that
$H(\gamma|\langle\alpha_1,\dots,\alpha_n\rangle)$ and $H(\gamma|\langle
\beta_1,\dots,\beta_n\rangle)$ are small. We say that ``extraction
of common information is impossible'' if the entropy of any such
variable $\gamma$ is small.

Let us show that this is the case if $\alpha$ and $\beta$ are
independent. In this case 
$\alpha^{n}=\langle\alpha_1,\ldots,\alpha_n\rangle$ and $\beta^{n}= \langle
\beta_1,\ldots,\beta_n\rangle$ are independent. Recall the well-known
inequality
        $$
H(\gamma) \le H(\gamma|\alpha^n)+H(\gamma|\beta^n)+I(\alpha^n:\beta^n).
        $$
Here $I(\alpha^n:\beta^n)=0$ (because $\alpha^n$ and $\beta^n$ are
independent); two other summands on the right hand side are small by
our assumption.

It turns out that a similar statement holds for dependent random
variables. However, there is one exception. If the joint probability
matrix of $\langle\alpha,\beta\rangle$ can be divided into blocks,
there is a random variable $\tau$ that is a function of 
$\alpha$ and a function of $\beta$ (``block number''). Then $\gamma =
\langle \tau_1, \dots, \tau_n\rangle$ is  common information of
$\alpha^n$ and $\beta^n$.

It was shown by Ahlswede, G\'acs and
K\"orner~\cite{ahlswedekorner,ahlswedekornerinternet,gacskorner}
that this is the only
case when there exists common information.

Their original proof is quite technical. Several years ago
another approach was proposed by Romashchenko~\cite{condind}
using ``conditionally independent'' random variables.
Romashchenko introduced the notion 
of conditionally independent random variables and showed
that extraction of common information from conditionally independent random
variables is impossible. We prove that if the joint probability matrix of a
pair of random variables $(\alpha, \beta)$ is not a block matrix, then 
$\alpha$ and $\beta$ are conditionally independent. We also show 
several new information inequalities for
conditionally independent random variables.

\section{Conditionally independent random
variables}

%%\begin{definition}
        %
Consider four random variables $\alpha,\beta,\alpha^*,\beta^*$.
Suppose that $\alpha^*$ and $\beta^*$ are independent, $\alpha$ and
$\beta$ are independent given $\alpha^*$, and also independent
given $\beta^*$, \textit{i.e.}, $I(\alpha^*:\beta^*)=0$,
$I(\alpha:\beta|\alpha^*)=0$ and $I(\alpha:\beta|\beta^*)=0$. Then
we say that $\alpha$ and $\beta$ are \emph{conditionally
independent of order $1$}. (Conditionally independent random
variables of order $0$ are independent random variables.)
        %
%%\end{definition}

We consider conditional independence of random variables as a
property of their joint distributions. If a pair of random variables $\alpha$ and $\beta$
has the same joint distribution as a pair of conditionally
independent random variables $\alpha_0$ and $\beta_0$ (on another probability space), we
say that $\alpha$ and $\beta$ are conditionally independent.

%%In other words,
%%random variables $\alpha$ and $\beta$ are conditionally
%%independent of order~$1$ if there exist random variables
%%$\gamma$ and $\delta$ in a subdivided probability space such
%%that the properties given above hold.

Replacing the requirement of independence of $\alpha^*$ and
$\beta^*$ by the requirement of conditional independence of order
$1$, we get the definition of conditionally independent random
variables ($\alpha$ and $\beta$) of order $2$ and so on.
(Conditionally independent variables of order $k$ are also called
$k$-conditionally independent in the sequel.)

\begin{definition}
We say that $\alpha$ and $\beta$ are \emph{conditionally independent
with respect to $\alpha^*$ and $\beta^*$} if $\alpha$ and $\beta$ are
independent given $\alpha^*$, and they are also independent given
$\beta^*$, \textit{i.e.} $I(\alpha:\beta|\alpha^*)=I(\alpha:\beta|\beta^*)=0$.
\end{definition}

\begin{definition} (Romashchenko~\cite{condind})
Two random variables $\alpha$ and $\beta$ are called conditionally independent random variables of order $k$ ($k\ge 0$)
if there exists a probability space $\Omega$ and a sequence of pairs of random variables 
$$\langle \alpha_0,\beta_0\rangle, \langle \alpha_1,\beta_1\rangle,\dots , \langle\alpha_k,\beta_k\rangle$$
on it such that 

(a) The pair $\langle\alpha_0, \beta_0\rangle$ has
the same distribution as $\langle \alpha,\beta\rangle$.

(b) $\alpha_i$ and $\beta_i$ are conditionally independent with respect to $\alpha_{i+1}$ and $\beta_{i+1}$ when $0\le i<k$.

%%        $$
%%I(\alpha_i:\beta_i|\alpha_{i+1})=0, \quad
%%I(\alpha_i:\beta_i|\beta_{i+1})=0,
%%        $$
%%, and $$I(\alpha_k:\beta_k)=0.$$

(c) $\alpha_k$ and $\beta_k$ are independent random variables.

The sequence 
   $$
\langle \alpha_0,\beta_0\rangle, \langle \alpha_1,\beta_1\rangle,\dots , \langle\alpha_k,\beta_k\rangle
   $$
is called a derivation for $\langle \alpha,\beta\rangle$.

We say that random variables $\alpha$ and $\beta$ are conditionally independent
if they are conditionally independent of some order $k$.
\end{definition}

%%Clearly, we can build all considered variables on
%%one probability space. Therefore, we can give an equivalent
%%definition for conditional independence:
%%a distribution $D$ of a pair of random variables is
%%conditionally independent of order $k$ if
%%there exist a probability space and random variables
%%        $$
%%\alpha_0,\beta_0, \alpha_1,\beta_1,\dots,\alpha_k,\beta_k
%%        $$
%%on it such that pair $\langle\alpha_0, \beta_0\rangle$ has
%%distribution $D$,
%%        $$
%%I(\alpha_i:\beta_i|\alpha_{i+1})=0, \quad
%%I(\alpha_i:\beta_i|\beta_{i+1})=0,
%%        $$
%%when $0\le i<k$, and $$I(\alpha_k:\beta_k)=0.$$

The notion of conditional independence can be applied for
analysis of common information using the following observations
(see below for proofs):

\begin{lemma} %%1.
Consider conditionally independent random variables $\alpha$ and
$\beta$ of order~$k$. 
Let $\alpha^n$ [$\beta^n$]
be a sequence of independent random variables each with the same distribution as 
$\alpha$ [$\beta$].
Then the variables $\alpha^n$ and $\beta^n$ are conditionally independent of order~$k$.
\end{lemma}

\begin {theorem} (Romashchenko~\cite{condind}) %%1.
If random variables $\alpha$ and $\beta$ are conditionally
independent of order~$k$, and $\gamma$~is an arbitrary random
variable (on the same probability space), then
        $$
H(\gamma) \le 2^k H(\gamma|\alpha)+ 2^k H(\gamma|\beta).
        $$
\end{theorem}

\smallskip

\begin{definition}
An $m\times n$ matrix is called a \emph{block matrix} if (after
some permutation of its rows and columns) it consists of four
blocks; the blocks on the diagonal are not equal to zero; the
blocks outside the diagonal are equal to zero.

Formally, $A$ is a block matrix if the set of its first indices
$\{1,\ldots,m\}$ can be divided into two disjoint nonempty sets
$I_1$ and $I_2$ ($I_1 \sqcup I_2 = \{1,\ldots,m\}$) and the set
of its second indices $\{1,\ldots,n\}$ can be divided into two
sets $J_1$ and $J_2$ ($J_1 \sqcup J_2 = \{1, \dots, n\}$) in
such a way that each of the blocks $\{a_{ij}: i \in I_1, j \in
J_1\}$ and $\{a_{ij}: i \in I_2, j \in J_2\}$ contains at least
one nonzero element, and all the elements outside these two
blocks are equal to $0$, \textit{i.e.} $a_{ij}=0$ when $(i,j) \in (I_1
\times J_2) \cup (I_2 \times J_1)$.
\end{definition}

\begin{theorem} %%2.
Random variables are conditionally independent \textit{iff} their joint
probability matrix is not a block matrix.
\end{theorem}

Using these statements, we conclude that if the joint
probability matrix of a pair of random variables
$\langle\alpha,\beta\rangle$ is not a block matrix, 
then no information
can be extracted from a sequence of $n$ independent
random variables each with the same distribution as 
$\langle \alpha ,\beta\rangle$:
    $$
H(\gamma) \le 2^k H(\gamma|\alpha^n)+ 2^k H(\gamma|\beta^n)
    $$
for some $k$ (that does not depend on $n$) and for any random
variable $\gamma$.
%%Thus we get another proof of one of the results
%%of G\'acs and K\"orner~\cite{gacskorner}.

\section{Proof of Theorem 1} %% change

%%Evidently, two random variables $\langle \alpha,\beta\rangle$ are
%%conditionally independent of order $k$ if and only if there exists
%%a derivation of length $k+1$ for $\langle \alpha,\beta\rangle$.

\noindent\hspace{2em}{\it Theorem 1: }
If random variables $\alpha$ and $\beta$ are conditionally
independent of order~$k$, and $\gamma$~is an arbitrary random
variable (on the same probability space), then
        $$
H(\gamma) \le 2^k H(\gamma|\alpha)+ 2^k H(\gamma|\beta).
        $$

\noindent\hspace{2em}{\it Proof
%% of Theorem 1
: }
The proof is by induction on $k$. The statement is already
proved for independent random variables $\alpha$ and $\beta$
($k=0$).

Suppose $\alpha$ and $\beta$ are conditionally independent with
respect to conditionally independent random variables $\alpha^*$
and $\beta^*$ of order $k-1$. From the conditional form of the
inequality
        $$
H(\gamma)  \le H(\gamma | \alpha) +
     H(\gamma | \beta) + I(\alpha:\beta)
        $$
($\alpha^*$ is added everywhere as a condition) it follows that
      \begin{multline*}
H(\gamma|\alpha^*) \le H(\gamma|\alpha \alpha^*) +
H(\gamma|\beta \alpha^*) + I(\alpha:\beta|\alpha^*)={}\\
H(\gamma|\alpha \alpha^*) + H(\gamma|\beta \alpha^*)  \le
H(\gamma|\alpha) +
H(\gamma|\beta).
        \end{multline*}
Similarly, $H(\gamma|\beta^*) \le H(\gamma|\alpha) +
H(\gamma|\beta)$. By the induction hypothesis $H(\gamma) \le
2^{n-1} H(\gamma|\alpha^*) + 2^{n-1} H(\gamma|\beta^*)$.
Replacing $H(\gamma|\alpha^*)$ and $H(\gamma|\beta^*)$ by their
upper bounds, we get $H(\gamma) \le 2^n H(\gamma|\alpha) + 2^n
H(\gamma|\beta)$.
\endproof

\begin {corollary} %% Corollary to Theorem 1
If the joint probability matrix $A$ of a pair of random
variables is a block matrix, then these random variables are not
conditionally independent.
\end{corollary}

\begin{proof}
Suppose that the joint probability matrix $A$ of random variables
$\langle\alpha, \beta\rangle$ is a block matrix and these random
variables are conditionally independent of order $k$.

Let us divide the matrix $A$ into blocks 
$I_1 \times J_1$ and $I_2 \times J_2$ as in Definition~3. Consider a
random variable $\gamma$ with two values that is equal to the
block number that contains $\langle\alpha,\beta\rangle$:
        \begin{align*}
\gamma = 1 &\Leftrightarrow \alpha\in I_1
           \Leftrightarrow \beta\in J_1;\\
\gamma = 2 &\Leftrightarrow \alpha\in I_2
           \Leftrightarrow \beta\in J_2.
        \end{align*}
The random variable $\gamma$ is a function of $\alpha$ and at
the same time a function of $\beta$. Therefore,
$H(\gamma|\alpha)=0$ and $H(\gamma|\beta)=0$. However, $\gamma$
takes two different values with positive probability. Hence
$H(\gamma)>0$, which contradicts Theorem~$1$.
\end{proof}

A similar argument shows that the order of conditional
independence should be large if the matrix is close to a block
matrix.

\section{Proof of Theorem 2}
For brevity, we call joint probability matrices of conditionally
independent random variables \emph{good matrices}.

The proof of Theorem 2 consists of three main steps. First, we
prove, that the set of good matrices is dense in the set of all
joint probability matrices. Then we prove that any matrix
without zero elements is good. Finally, we consider the general case
and prove that any matrix that is not a block matrix is good.

The following statements are used in the sequel.

(a)~The joint probability matrix of independent random
variables is a matrix of rank $1$ and vice versa.
In particular, all matrices of rank $1$ are good.

(b)~If $\alpha$ and $\beta$ are conditionally independent, $\alpha'$ is
a function of $\alpha$ and $\beta'$ is a function of $\beta$, then
$\alpha'$ and $\beta'$ are conditionally independent. (Indeed, if
$\alpha$ and $\beta$ are conditionally independent with respect to
some $\alpha^*$ and $\beta^*$, then $\alpha'$ and $\beta'$ are also
conditionally independent with respect to $\alpha^*$ and $\beta^*$.)

(c)~If two random variables are $k$-conditionally independent,
then they are $l$-conditionally independent for any~$l>k$. (We
can add some constant random variables to the end of the
derivation.)

(d)~Assume that conditionally independent random variables
$\alpha_1$ and $\beta_1$ are defined on a probability space
$\Omega_1$ and conditionally independent random variables
$\alpha_2$ and $\beta_2$ are defined on a probability space
$\Omega_2$. Consider random variables $(\alpha_1,\alpha_2)$ and
$(\beta_1,\beta_2)$ that are defined in a natural way on the
Cartesian product $\Omega_1\times\Omega_2$. Then $(\alpha_1,\alpha_2)$
and $(\beta_1,\beta_2)$ are conditionally independent.
Indeed, for each pair $\langle\alpha_i,\beta_i\rangle$ consider its
derivation
       $$
\langle\alpha_i^0,\beta_i^0\rangle,
\langle\alpha_i^1,\beta_i^1\rangle,\dots,
\langle\alpha_i^l,\beta_i^l\rangle
        $$
(using (c), we may assume that both derivations have the same
length~$l$).

Then the sequence
        $$
\langle(\alpha_1^0,\alpha_2^0),(\beta_1^0,\beta_2^0)\rangle,\ldots,
\langle(\alpha_1^l,\alpha_2^l),(\beta_1^l,\beta_2^l)\rangle
        $$       
is a derivation for the pair of random variables 
$\langle (\alpha_1,\alpha_2),(\beta_1,\beta_2)\rangle$. 
For example, random
variables $(\alpha_1,\alpha_2)=(\alpha_1^0,\alpha_2^0)$ and
$(\beta_1,\beta_2)=(\beta_1^0,\beta_2^0)$ are independent given the
value of $(\alpha_1^1,\alpha_2^1)$, because $\alpha_1$ and $\beta_1$ are
independent given $\alpha_1^1$, variables $\alpha_2$ and $\beta_2$ are
independent given $\alpha_2^1$, and the measure on
$\Omega_1\times\Omega_2$ is equal to the product of the measures on
$\Omega_1$ and $\Omega_2$.

Applying~(d) several times, we get Lemma~1.

Combining Lemma~1 and~(b), we get the following statement:

(e)~Let $\langle\alpha_1,\beta_1\rangle, \ldots,
\langle\alpha_n,\beta_n\rangle$ be independent and identically
distributed random variables. Assume that the variables in each
pair $\langle \alpha_i,\beta_i\rangle$ are conditionally
independent. Then any random variables $\alpha'$ and $\beta'$, where
$\alpha'$ depends only on $\alpha_1,\ldots,\alpha_n$ and $\beta'$ depends
only on $\beta_1,\ldots,\beta_n$, are conditionally independent.

\begin{definition}
Let us introduce the following notation:
        $$
D_\varepsilon=
\mm{1/2-\varepsilon}{\varepsilon}{\varepsilon}{1/2-\varepsilon}
        $$
(where $0\le \varepsilon\le 1/2$).
\end{definition}

The matrix $D_{1/4}$ corresponds to a pair of independent random
bits; as $\varepsilon$ tends to $0$ these bits become more
dependent (though each is still uniformly distributed over
$\{0,1\}$).

\smallskip

\begin{lemma} %%2.

%%\noindent
(i)~$D_{1/4}$ is a good matrix.

%%\noindent
(ii)~If $D_\varepsilon$ is a good matrix then
$D_{\varepsilon(1-\varepsilon)}$ is good.

%%\noindent
(iii)~There exists an arbitrary small $\varepsilon$ such that
$D_\varepsilon$ is good.
\end{lemma}

\smallskip

\begin{proof}

(i)~The matrix $D_{1/4}$ is of rank $1$, hence it is good
(independent random bits).

(ii)~Consider a pair of random variables $\alpha$ and $\beta$
distributed according to $D_\varepsilon$.

Define new random variables $\alpha'$ and~$\beta'$ as follows:

\begin{itemize}
        \item
if $(\alpha,\beta) = (0,0)$ then $(\alpha',\beta') = (0,0)$;
        \item
if $(\alpha,\beta) = (1,1)$ then $(\alpha',\beta') = (1,1)$;
        \item
if $(\alpha,\beta) = (0,1)$ or  $(\alpha,\beta) = (1,0)$ then

 $$(\alpha',\beta') =
\left\{
 \begin{array}{rl}
  (0,0) & \mbox{with probability } \varepsilon/2;     \\
  (0,1) & \mbox{with probability } (1-\varepsilon)/2; \\
  (1,0) & \mbox{with probability } (1-\varepsilon)/2; \\
  (1,1) & \mbox{with probability } \varepsilon/2.
 \end{array}
\right.
$$
\end{itemize}

The joint probability matrix of $\alpha'$ and $\beta'$ given $\alpha=0$
is equal to
       $$
\mm{(1-\varepsilon)^2}{\varepsilon(1-\varepsilon)}
{\varepsilon(1-\varepsilon )}{\varepsilon^2}
       $$
and its rank equals $1$. Therefore, $\alpha'$ and $\beta'$ are
independent given $\alpha=0$.

Similarly, the joint probability matrix of $\alpha'$ and $\beta'$ given $\alpha=1$, 
$\beta = 0$ or $\beta = 1$
has rank $1$. This yields that $\alpha'$
and $\beta'$ are conditionally independent with respect to $\alpha$
and $\beta$, hence $\alpha'$ and $\beta'$ are conditionally
independent.

The joint distribution of $\alpha'$ and $\beta'$ is
        $$
\mm{1/2-\varepsilon(1-\varepsilon)}{\varepsilon(1-\varepsilon)}
{\varepsilon(1-\varepsilon)}{1/2-\varepsilon(1-\varepsilon)},
        $$
hence~$D_{\varepsilon(1-\varepsilon)}$~is a good matrix.

(iii)~Consider the sequence $\varepsilon_n$ defined by
$\varepsilon_0=1/4$ and $\varepsilon_{n+1}=\varepsilon_n
(1-\varepsilon_n)$. The sequence $\varepsilon_n$ tends to zero
(its limit is a root of the equation $x=x(1-x)$). It follows
from statements (i) and~(ii) that all matrices
$D_{\varepsilon_n}$ are good.
\end{proof}

\note
The order of conditional independence of
$D_\varepsilon$ tends to infinity as $\varepsilon\to0$.
Indeed, applying Theorem 1 to random variables $\alpha$ and $\beta$ with
joint distribution $D_\varepsilon$ and  to $\gamma = \alpha$, we obtain
       $$
H(\alpha) \le 2^k
(H(\alpha|\alpha) + H(\alpha | \beta)) = 2^k H(\alpha | \beta).
        $$
Here $H(\alpha)=1$; for any fixed value of $\beta$
the random variable $\alpha$ takes two values with probabilities
$2\varepsilon$ and  $1-2\varepsilon$, therefore
        $$
H(\alpha|\beta) = -(1-2 \varepsilon) \log_2 (1 - 2 \varepsilon) -
2\varepsilon \log_2 (2\varepsilon) = O(-\varepsilon \log_2 \varepsilon)
        $$
and  (if $D_\varepsilon$ corresponds to conditionally
independent variables of order $k$)
        $$
2^k \ge H(\alpha)/H(\alpha|\beta) = 1/O(-\varepsilon \log_2
\varepsilon) \to \infty
        $$
as $\varepsilon \to 0$.

\begin{lemma} %%3.
The set of good matrices is dense in the set of all joint
probability matrices (\textit{i.e.}, the set of $m\times n$ matrices with
non-negative elements, whose sum is $1$).
\end{lemma}

\begin{proof}
Any joint probability matrix $A$ can be approximated as closely as desired by matrices
with elements of the form $l/2^N$ for some $N$ (where $N$ is the
same for all matrix elements).

Therefore, it suffices to prove that any joint probability
matrix $B$ with elements of the form $l/2^N$  can be approximated (as closely as desired) by good
matrices. Take a pair of random variables $\langle \alpha, \beta \rangle$ distributed according to $D$.
The pair $\langle \alpha, \beta \rangle$ can be represented as a function of 
$N$ independent Bernoulli trials. The joint distribution matrix of each of these
trials is $D_0$ and, by Lemma~2, can be approximated by a good matrix. 
Using statement (e), we get that $\langle \alpha, \beta \rangle$ can also be approximated by a good matrix.
Hence $D$ can be approximated as closely as desired by good matrices.
\end{proof}

\begin{lemma} %%4.
 If $A=(a)_{ij}$ and $B=(b)_{ij}$ are stochastic matrices and $M$ is a good matrix, then $A^TMB$ is a good matrix.
\end{lemma}
\begin{proof}
Consider a pair of random variables $\langle \alpha, \beta \rangle$
distributed according to $M$. This pair of random variables is conditionally independent.

Roughly speaking, we define random variable $\alpha'$ [$\beta'$] as a transition from $\alpha$ [$\beta$]
with transition matrix $A$ [$B$]. The joint probability matrix of $\langle\alpha',\beta'\rangle$ is equal to $A^TMB$.
But since the transitions are independent from $\alpha$ and $\beta$, the new random variables are conditionally
independent.

More formally, let us randomly (independently from $\alpha$ and $\beta$)
choose vectors $\vec{c}$ and $\vec{d}$ as follows
$$\Pr (\proj_i (\vec{c}) = j) = a_{ij},$$
$$\Pr (\proj_i (\vec{d}) = j) = b_{ij},$$
where $\proj_i$ is the projection onto the $i$-th component. 

Define $\alpha' = \proj_{\alpha} (\vec{c})$ and 
$\beta' = \proj_{\beta} (\vec{d})$. Then 

(i)~the joint probability matrix of $\langle\alpha',\beta'\rangle$ is equal to $A^TMB$;

(ii)~the pair $(\alpha, \vec{c})$ is conditionally independent from the pair $(\beta, \vec{d})$. Hence
by statement~(b), $\alpha'$ and $\beta'$ are conditionally independent. 

\end{proof}

Now let us prove the following technical lemma.

\begin{lemma} %%5
For any nonsingular $n \times n$ matrix $M$ and  a matrix $R = (r)_{ij}$ with the sum of its
elements equal to 0, there exist matrices $P$ and $Q$ such that 

1. $R =  P^TM + MQ$;

2. the sum of all elements in each row of $P$ is equal to $0$;

3. the sum of all elements in each row of $Q$ is equal to $0$.
\end{lemma}

\begin{proof}
First, we assume that $M = I$
(here $I$ is the identity matrix of the proper size),
and find matrices $P'$ and $Q'$ such that

$$R = P'^T + Q'.$$

Let us define $P'=(p')_{ij}$ and $Q'=(q')_{ij}$ as follows:

$$q'_{ij} = \frac{1}{n} \sum_{k=1}^{n} r_{kj}.$$
Note that all rows of $Q'$ are the same and equal to the average of rows of $R$.

$$P' = (R - Q')^T$$

It is easy to see that condition (1) holds. Condition (3) holds because the sum of all elements
in any row of $Q$ is equal to the sum of all elements of $R$ divided by $n$, which is $0$ by
the condition. Condition (2) holds because
	$$
\sum_{j=1}^{n} p'_{ij} = \sum_{j=1}^{n} \left( r_{ji} - \frac{1}{n} \sum_{k=1}^{n} r_{ki} \right) = 0.
	$$

Now we consider the general case. Put $P = (M^{-1})^T P'$ and  $Q = M^{-1}Q'$.
Clearly (1) holds. Conditions (2) and (3) can be rewritten as
$P\vec{u}=0$ and $Q\vec{u}=0$, where $\vec{u}$ is the vector consisting of ones. But
$P\vec{u} = (M^{-1})^T (P'\vec{u}) = 0$ and $Q\vec{u} = M^{-1}(Q'\vec{u}) = 0$. Hence (2) and (3) hold.
\end{proof}

By altering the signs of $P$ and $Q$ we get Corollary~5.1.
 
\begin{corollary}
 For any nonsingular matrix $M$ and  a matrix $R$ with the sum of its
elements equal to 0, there exist matrices $P$ and $Q$ such that 

1. $R =  -P^TM - MQ$;

2. the sum of all elements in each row of $P$ is equal to $0$;

3. the sum of all elements in each row of $Q$ is equal to $0$.
\end{corollary}

\begin{lemma}
Any nonsingular matrix $M$ without zero elements is good.
\end{lemma}
 
\begin{proof}
Let $M$ be a nonsingular $n\times n$ matrix without zero elements. By Lemma~4, it suffices to show that $M$ can be represented as
    $$
 M = A^TGB,
    $$
where $G$ is a good matrix; $A$ and $B$ are stochastic matrices.
In other words, we need to find invertible stochastic matrices $A$, $B$ such that
$(A^T)^{-1}MB^{-1}$ is a good matrix. 

Let $V$ be the affine space of all $n\times n$ matrices in which the sum of
all the elements is equal to $1$:

$$V = \{ X:\sum_{i=1}^n\sum_{j=1}^n x_{ij} = 1\}.$$
(This space contains the set of all joint probability matrices.)

Let $U$ be the affine space of all $n \times n$ matrices in which the sum of
all elements in each row is equal to $1$:

$$U = \{ X:\sum_{j=1}^n x_{ij} = 1 \mbox{ for all } i \}.$$
(This space contains the set of stochastic matrices.)

Let $\tilde{U}$ be a neighborhood of $I$ in $U$ such that all matrices from this neighborhood
are invertible. Define a mapping $\psi:\tilde{U}\times\tilde{U} \rightarrow V $ as follows:
   $$
\psi (A, B) = (A^T)^{-1}MB^{-1}.
   $$  
  
Let us show that the differential of this mapping at the point
$A=B=I$ is a surjective mapping from $T_{(I,I)}\tilde{U}\times\tilde{U}$ (the tangent space of $\tilde{U}\times\tilde{U}$ at the point $(I, I)$) to $T_MV$ (the tangent space of $V$ at the point $M$).
Differentiate $\psi$ at $(I, I)$:
   $$
d\psi|_{A=I,\ B=I} = d\left((A^T)^{-1}MB^{-1}\right)= - (dA)^TM - M dB.
   $$
  
We need to show that for any matrix $R \in T_MV$, there exist matrices
$(P,Q) \in T_{(I,I)}\tilde{U}\times \tilde{U}$ such that

$$R = - P^TM - MQ.$$
But this is guaranteed by Corollary~5.1.

Since the mapping $\varphi$ has a surjective differential at $(I,I)$,
it has a surjective differential in some neighborhood $N_1$ of  $(I,I)$ in $\tilde{U}\times \tilde{U}$.
Take a pair of stochastic matrices $(A_0, B_0)$ from this neighborhood such that
these matrices are interior points of the set of stochastic matrices.

Now take a small neighborhood $N_2$ of $(A_0, B_0)$ from the intersection of $N_1$ and the set of
stochastic matrices. Since the differential of $\varphi$ at $(A_0, B_0)$ is surjective, the image of $N_2$
has an interior point. Hence it contains a good matrix
(recall that the set of good matrices is dense in the set of all joint probability matrices).
In other words, $\psi (A_1, B_1) = (A_1^T)^{-1}MB_1^{-1}$ is a good matrix for some pair of
stochastic matrices $(A_1, B_1) \in N_2$.
This finishes the proof.
\end{proof}

\begin{lemma} %%7.
Any joint probability matrix without zero elements is a good
matrix.
\end{lemma}

\begin{proof}
Suppose that $X=(\vec v_1, \dots \vec v_n)$ is an  $m\times n$ ($m > n$) matrix
of rank $n$.
%Let $X=(\vec v_1, \dots \vec v_n)$ be an  $m \times n$ ($m > n$) matrix
%of rank $n$.
 It is equal to  the product of a nonsingular matrix and
stochastic matrix:
        \begin{multline*}
\textstyle
X=(\vec v_1 - \vec u_1 - \ldots - \vec u_{m-n},\vec v_2, \dots, \vec v_n, \vec
u_1, \dots, \vec u_{m-n})\times{}\\
{}\times\left(
\arraycolsep=3pt
\begin{array}{cccc}
\multicolumn{4}{c}{\displaystyle I} \\[-3pt]
\scriptstyle 1 & \scriptstyle 0 & \scriptstyle \ldots & \scriptstyle 0 \\[-3pt]
\scriptstyle \vdots & \scriptstyle \vdots & \scriptstyle \ddots & \scriptstyle \vdots\\[-3pt]
\scriptstyle 1 & \scriptstyle 0 & \scriptstyle \ldots &\scriptstyle 0
\end{array}
\right)
        \end{multline*}
%%\left( \begin{array}{cccc} 1 & 0 & \ldots & 0 %%\\ 0 & 1 & \ldots & 0
%%\\ \vdots & \vdots & \ddots & \vdots \\ 0 & 0 & \ldots & 1 \\
%%1 & 0 & \ldots & 0 \\ \vdots & \vdots & \vdots & \vdots
%%\\1 & 0 & \ldots & 0 \end{array} \right)
        %
where $\vec u_1, \dots, \vec u_{m-n}$ are sufficiently small
vectors with positive components that form a basis in
${\mathbb R}^m$ together with $\vec v_1,\dots, \vec v_n$ (it is
easy to see that such vectors do exist); vectors $\vec u_1,
\dots, \vec u_{m-n}$ should be small enough to ensure that
the vector $\vec v_1 - \vec u_1 - \ldots - \vec u_{m-n}$ has positive elements.

The first factor is a nonsingular matrix with positive
elements and hence is good. The second factor is a stochastic
matrix, so the product is a good matrix.

Therefore, any matrix of full rank without zero elements is
good. If a $m\times n$ matrix with positive
elements does not have full rank, we can add (in a similar way)
$m$ linearly independent columns to get a matrix of full
rank and then represent the given matrix as a product
of a matrix of full rank and stochastic matrix.
\end{proof}

We denote by $S(M)$ the sum of all elements of a matrix $M$.

\begin{lemma} %%8.
Consider a matrix $N$ whose elements are matrices
$N_{ij}$ of the same size. If

(a)~all $N_{ij}$ contain only nonnegative elements;

(b)~the sum of matrices in each row and in each column of the
matrix $N$ is a matrix of rank $1$;

(c)~the matrix $P$ with elements
$p_{ij}=S(N_{ij})$ is a good joint probability matrix;

\noindent%%end of condition (c)
then the sum of all the matrices  $N_{ij}$ is a good matrix.
\end{lemma}

\begin{proof}
This lemma is a reformulation of the definition of conditionally
independent random variables. Consider random variables $\alpha^*,
\beta^*$ such that the probability of the event $\langle \alpha^*,
\beta^*\rangle = \langle i,j\rangle$ is equal to $p_{ij}$, and
the probability of the event
        $$
\alpha=k,\ \beta=l,\ \alpha^*=i,\ \beta^*=j
        $$
is equal to the $(k,l)$-th element of the matrix $N_{ij}$.

The sum of matrices $N_{ij}$ in a row $i$ corresponds to the
distribution of the pair $\langle \alpha,\beta\rangle$ given
$\alpha^*=i$; the sum of matrices $N_{ij}$ in a column $j$
corresponds to the distribution of the pair $\langle
\alpha,\beta\rangle$ given $\beta^*=j$; the sum of all the matrices
$N_{ij}$ corresponds to the distribution of the pair $\langle
\alpha,\beta\rangle$.
\end{proof}

From Lemma~8 it follows that any $2\times2$ matrix of the form
$\mm{a}{b}{0}{c}$ is good.%%
\footnote{$a$, $b$ and $c$ are positive numbers whose sum equals~$1$.}
Indeed, let us apply Lemma~8 to the following matrix:
        $$
N=\mmmm{\mmm{a}{0}{0}{0}}
   {\mmm{0}{b/2}{0}{0}}
   {\mmm{0}{b/2}{0}{0}}
   {\mmm{0}{0}{0}{c}}.
        $$
The sum of matrices in each row and in each column is of
rank~$1$. The sum of elements of each matrix $N_{ij}$ is
positive, so (by Lemma~7) the matrix $p_{ij} = S(N_{ij})$ is a
good matrix. Hence the sum of matrices $N_{ij}$ is good.

Recalling that 
%%$*$-symbols 
$a$, $b$ and $c$ 
stand for any positive numbers whose
sum is~$1$, we conclude that any $2\times2$-matrix with $0$ in
the left bottom corner and positive elements elsewhere is a good
matrix. Combining this result with the result of Lemma~7, we 
get that any non-block $2\times 2$ matrix is good.

In the general case (we have to prove that any non-block matrix
is good) the proof is more complicated.

We will use the following definitions:

\begin{definition}
{The \emph{support} of a matrix is the set of positions of its
nonzero elements.}
An \emph{r-matrix} is a matrix with
nonnegative elements and with a ``rectangular'' support (\textit{i.e.},
with support $A\times B$ where $A$[$B$] is some set of rows[columns]).
\end{definition}

\begin{lemma} %%9
Any  r-matrix~$M$ is the sum of some r-matrices of rank~1 with the
same support as~$M$.
\end{lemma}

\begin{proof}
Denote the support of $M$ by $N=A\times B$. Consider the basis
$E_{ij}$ in the vector space of matrices whose support is a subset
of~$N$. (Here $E_{ij}$ is the matrix that has $1$ in
the $(i,j)$-position and $0$ elsewhere.)

The matrix $M$ has positive coordinates in the basis $E_{ij}$. 
Let us approximate each matrix $E_{ij}$ by a slightly different matrix
$E'_{ij}$
of rank~$1$ with support $N$:
$$
E'_{ij} = \left(\vec e_i + \varepsilon \sum_{k \in A} \vec e_k\right) \cdot
          \left(\vec e_j + \varepsilon \sum_{l \in B} \vec e_l\right)^T,
$$
where $\vec e_1, \dots, \vec e_n$ is the standard basis in ${\mathbb R}^n$.

 The coordinates $c_{ij}$ of $M$ in the new basis $E'_{ij}$ continuously
depend on $\varepsilon$. Thus they remain positive if $\varepsilon$ is sufficiently small.
So taking a sufficiently small $\varepsilon$
we get the required representation of $M$ as the sum of matrices of rank~$1$ with support $N$:
 
 $$M = \sum_{(i,j) \in N} c_{ij} E'_{ij}.$$
\end{proof}

\begin{definition}
An \emph{r-decomposition} of a matrix is its expression as a
(finite) sum of r-matrices $M=M_1+M_2+\ldots$ of the same size
such that the supports of $M_i$ and $M_{i+1}$ intersect (for any
$i$). The \emph{length} of the decomposition is the number of the
summands; the \emph{r-complexity} of a matrix is the length of
its shortest decomposition (or $+\infty$, if there is no
such decomposition).
\end{definition}

\begin{lemma} %%9.
Any non-block matrix $M$ with nonnegative elements has an
r-decomposition.
\end{lemma}

\begin{proof}
Consider a graph whose vertices are nonzero entries of $M$. Two
vertices are connected by an edge \textit{iff} they are in the same row
or column. By assumption, the matrix is a non-block matrix, hence the
graph is connected and there exists a (possibly non-simple) path $(i_1,j_1) \dots
(i_m,j_m)$ that visits each vertex of the graph at least once.

Express $M$ as the sum of matrices corresponding to the edges of
the path: each edge corresponds to a matrix whose support
consists of the endpoints of the edge; each positive element of $M$ 
is distributed among matrices corresponding to the adjacent edges.
%%We distribute each positive element of $M$ among matrices
%%corresponding to the adjacent edges.
Each of these matrices is of rank $1$. So the expression 
of $M$ as the sum of these matrices is an r-decomposition.

\end{proof}

\begin{corollary} %%corollary to Lemma 8
The r-complexity of any non-block matrix is finite.
\end{corollary}

\smallskip

\begin{lemma} %%11.
Any non-block matrix $M$ is good.
\end{lemma}

\smallskip

\begin{proof}
The proof uses induction on r-complexity of $M$. For matrices of
r-complexity $1$, we apply Lemma~7.

Now suppose that $M$ has r-complexity~$2$. In this case $M$ is
equal to the sum of some r-matrices $A$ and $B$ such that their
supports are intersecting rectangles. By Lemma~9, each of the
matrices $A$ and $B$ is the sum of matrices of rank~1 with the
same support.

Suppose, for example, that $A=A_1+A_2+A_3$ and $B=B_1+B_2$.
Consider the block matrix
        $$
\left(
 \begin{array}{ccccc}
     A_1 & 0 & 0 & 0 & 0 \\
     0 & A_2 & 0 & 0 & 0 \\
     0 & 0 & A_3 & 0 & 0 \\
     0 & 0 & 0 & B_1 & 0 \\
     0 & 0 & 0 & 0 & B_2 \\
 \end{array}
\right).
        $$
The sum of the matrices in each row and in each column is a
matrix of rank~1. The sum of all the entries is equal to $A+B$.
All the conditions of Lemma~8 but one hold. The only problem
is that the matrix $p_{ij}$ is diagonal and hence is not
good, where $p_{ij}$ is the sum of the elements of the matrix
in the $(i,j)$-th entry (see Lemma~8). To overcome this
obstacle take a matrix $e$ with only one nonzero element that is
located in the intersection of the supports of $A$ and $B$.
If this nonzero element is sufficiently small, then all the
elements of the matrix
        $$
\arraycolsep=0.7\arraycolsep
N=\left(
 \begin{array}{ccccc}
     A_1 - 4e  & e & e & e & e \\
     e & A_2 -4e & e & e & e \\
     e & e & A_3 -4e & e & e \\
     e & e & e & B_1 -4e & e \\
     e & e & e & e & B_2 -4e \\
 \end{array}
\right)
        $$
are nonnegative matrices. The sum of the elements of each of the matrices
that form the matrix $N$ is positive. And the sum of the elements in
any row and in any column is not changed, so it is of rank~$1$.
Using Lemma~8 we conclude that the matrix $M$ is good.

The proof for matrices of r-complexity~$3$ is similar. For
simplicity, consider the case where a matrix of complexity~3
has an r-decomposition $M=A+B+C$, where $A, B, C$ are r-matrices
of rank $1$. Let $e_1$ be a matrix with one positive element
that belongs to the intersection of the supports of $A$ and $B$
(all other matrix elements are zeros), and $e_2$ be a matrix
with a positive element in the intersection of the supports
of~$B$ and $C$.

Now consider the block matrix
        $$
N=\left(
 \begin{array}{ccc}
     A - e_1 & e_1 & 0 \\
     e_1 & B-e_1-e_2 & e_2 \\
     0  & e_2 & C-e_2 \\
 \end{array}
\right).
        $$
Clearly, the sums of the matrices in each row and in each column
are of rank~$1$. The support of the matrix $(p)_{ij}$ is of the form
        $$
\left(
 \begin{array}{ccc}
     * & *  & 0 \\
     * & *  & * \\
     0 & *  & * \\
 \end{array}
\right);
        $$
and $(p)_{ij}$ has r-complexity~$2$.\footnote{Its support is the union of
two intersecting rectangles, so the matrix is the sum of two
r-matrices.} By the inductive assumption any matrix of
r-complexity~$2$ is good. Therefore, $M$ is a good matrix
(Lemma~8).

In the general case (any matrix of r-complexity~$3$) the reasoning
is similar. Each of the matrices $A,B,C$ is represented as the sum
of some matrices of rank~1 (by Lemma~9). Then we need several
entries $e_1$ ($e_2$) (as it was for matrices of
r-complexity~$2$).
In the same way, we prove the lemma for matrices of
r-complexity~$4$ \textit{etc}.
\end{proof}

This concludes the proof of Theorem~2: Random variables are
conditionally independent if and only if their joint probability
matrix is a non-block matrix.

Note that this proof is ``constructive'' in the following sense.
Assume that the joint probability matrix for $\alpha,\beta$ is given
and this matrix is not a block matrix. (For simplicity we assume
that matrix elements are rational numbers, though this is not an
important restriction.) Then we can effectively find $k$ such that $\alpha$ and $\beta$ are
$k$-independent, and find the joint distribution of all random
variables that appear in the definition of $k$-conditional
independence. (Probabilities for that distribution are not
necessarily rational numbers, but we can provide algorithms that
compute approximations with arbitrary precision.)

\section{Improved version of Theorem~1}

The inequality
$$ H(\gamma) \le   2^k H(\gamma|\alpha) +  2^k H(\gamma|\beta)   $$
from Theorem 1 can be improved. In this section we prove  a stronger
theorem.

\begin{theorem} %%3.
 If random variables $\alpha$ and $\beta$
are  conditionally independent of order~$k$, and $\gamma$~is
an arbitrary random variable, then
        $$
 H(\gamma) \le   2^k H(\gamma|\alpha) +  2^k H(\gamma|\beta)
-(2^{k+1} -1) H(\gamma|\alpha \beta),
        $$
or, in another form,
        $$
I(\gamma:\alpha\beta) \le 2^k I(\gamma:\alpha|\beta) + 2^k
I(\gamma:\beta|\alpha).
        $$
\end{theorem}

\medskip

\begin{proof}
 The proof is by induction on $k$.

We use the following inequality:
        \begin{multline*}
H(\gamma) = H(\gamma|\alpha) + H(\gamma|\beta)  + \\I(\alpha:\beta)-
I(\alpha:\beta|\gamma) - H(\gamma |  \alpha\beta) \le\\
H(\gamma|\alpha) + H(\gamma|\beta)  + I(\alpha:\beta) - H(\gamma |
\alpha\beta).
        \end{multline*}
If  $\alpha$ and $\beta$~are independent then $I(\alpha:\beta) = 0$,
we get the required inequality.

Assume that $\alpha$ and $\beta$ are conditionally
independent with respect to $\alpha'$ and $\beta'$;
$\alpha'$ and $\beta'$~are conditionally independent of
order $k-1$.

We can assume without loss of generality that two random
variables, the pair $\langle\alpha', \beta'\rangle$, and $\gamma$
are independent given $\langle\alpha, \beta\rangle$. Indeed,
consider random variables $\langle\alpha^*, \beta^*\rangle$
defined by the following formula
        \begin{multline*}
\Pr(\alpha^*=c, \beta^*=d |\alpha=a, \beta=b, \gamma = g ) =\\
\Pr(\alpha'=c, \beta'=d |\alpha=a,
\beta=b).
        \end{multline*}
The distribution of $\langle\alpha, \beta,
\alpha^*,\beta^*\rangle$ is the same as the distribution of
$\langle\alpha, \beta, \alpha',\beta'\rangle$, and
$\langle\alpha^*,\beta^*\rangle$ is independent from $\gamma$
given $\langle\alpha, \beta\rangle$.

From the ``relativized'' form of the inequality
        $$
H(\gamma) \le H(\gamma|\alpha) + H(\gamma|\beta) +
I(\alpha:\beta) - H(\gamma | \alpha\beta)
        $$
($\alpha'$ is added as a condition everywhere) it follows that
        \begin{multline*}
H(\gamma|\alpha') \le\\ H(\gamma|\alpha\alpha') +
H(\gamma|\beta\alpha') + I(\alpha:\beta|\alpha') - H(\gamma |
\alpha' \alpha\beta)\le\\ H(\gamma|\alpha) + H(\gamma|\beta) -
H(\gamma | \alpha' \alpha\beta).
        \end{multline*}
Note that according to our assumption $\alpha'$ and $\gamma$ are
independent given $\alpha$ and $\beta$, so $H(\gamma | \alpha'
\alpha\beta) = H(\gamma | \alpha\beta)$.

Using the upper bound for $H(\gamma | \alpha')$, the similar bound
for $H(\gamma | \beta')$ and the induction assumption, we
conclude that
\begin{multline*}
H(\gamma) \le 2^k H(\gamma|\alpha) + 2^k H(\gamma|\beta)\\ - 2^k
H(\gamma|\alpha \beta) - (2^k -1) H(\gamma|\alpha' \beta').
\end{multline*}
Applying the inequality 
$$ H(\gamma|\alpha' \beta') \ge H(\gamma
| \alpha' \beta' \alpha \beta) = H(\gamma|\alpha \beta),$$ we
get the statement of the theorem.
\end{proof}

\section{Rate Regions}

\begin{definition}
The rate region of a pair of random variables $\alpha,\beta$ is the
set of triples of real numbers $\langle u, v, w\rangle$ such
that for all $\varepsilon >0$, $\delta >0$ and sufficiently
large $n$ there exist
        \begin{itemize}
\item ``coding'' functions $t$, $f$ and $g$; their arguments are
pairs $\langle\alpha^n,\beta^n\rangle$; their values are binary
strings of length $\lfloor (u + \delta) n\rfloor$, $\lfloor
(v+\delta) n\rfloor$ and $\lfloor (w+\delta) n\rfloor$
(respectively).
\item ``decoding'' functions $r$ and $s$ such that
$$r(t(\alpha^n,\beta^n), f(\alpha^n,\beta^n))=\alpha^n$$
and
$$s(t(\alpha^n,\beta^n), g(\alpha^n,\beta^n))=\beta^n$$ with probability
more then $1 - \varepsilon$.
        \end{itemize}
\end{definition}

\begin{figure}
        $$
\includegraphics{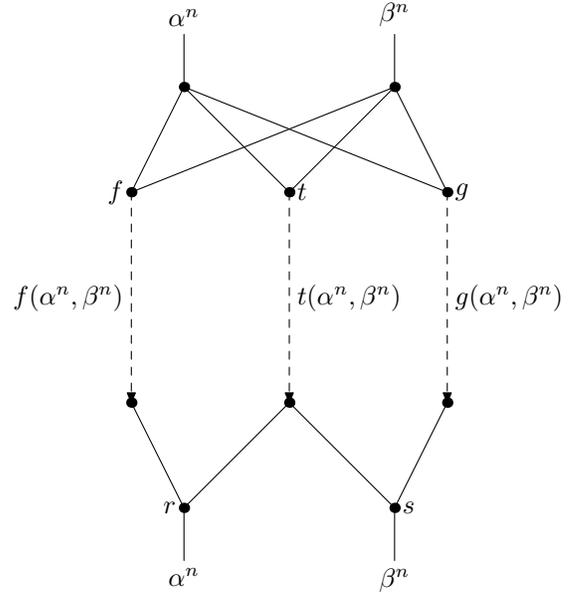}
%%epsffile{ieee1.eps}
%%\epsfbox{ieee1.eps}
        $$
\caption{Values of $\alpha^n$ and $\beta^n$ are encoded by
functions $f$, $t$ and $g$ and then transmitted via
channels of limited capacity (dashed lines);
decoder functions $r$ and $s$ have to reconstruct
values $\alpha^n$ and $\beta^n$ with high probability
having access only to a part of transmitted information.}
\label{multisource}
\end{figure}

This definition (standard for multisource coding theory, see \cite{csizarkorner})
corresponds to the scheme of information transmission
presented on Figure~\ref{multisource}.

The following theorem was discovered by Vereshchagin. It gives a new constraint 
on the rate region when $\alpha$ and $\beta$ are conditionally independent.

\begin{theorem} %% 4.
Let $\alpha$ and $\beta$ be
$k$-conditionally independent random variables.
Then,
  $$H(\alpha)+H(\beta)\le v+w+(2-2^{-k})u$$
for any triple $\langle u, v, w\rangle$ in the rate region.
\end{theorem}

(It is easy to see that $H(\alpha)\le u+v$ since $\alpha^n$ can be
reconstructed with high probability from strings of length
approximately $nu$ and $nv$. For similar reasons we have
$H(\beta)\le u+w$. Therefore,
        $$
H(\alpha)+H(\beta)\le v+w+2u
        $$
for any $\alpha$ and $\beta$. Theorem~4 gives a stronger bound for
the case when $\alpha$ and $\beta$ are $k$-independent.)

\begin{proof}
Consider random variables
        $$
\gamma = t(\alpha^n,\beta^n), \xi = f(\alpha^n,\beta^n),
\eta = g(\alpha^n,\beta^n)
        $$
from the definition of the rate region (for some fixed
$\varepsilon>0$). By Theorem~1, we have
        $$
H(\gamma) \le 2^k(H(\gamma|\alpha^n) +
H(\gamma|\beta^n)).
        $$
We can rewrite this inequality as
        $$
2^{-k} H(\gamma) \le H(\langle\gamma,\alpha^n\rangle) +
H(\langle\gamma,\beta^n\rangle) - H(\alpha^n) -H(\beta^n)
        $$
or
        \begin{multline*}
H(\xi) + H(\eta) + (2-2^{-k}) H(\gamma) \ge H(\xi) +
H(\eta) +\\ 2 H(\gamma)-H(\langle\gamma,\alpha^n\rangle) -
H(\langle\gamma,\beta^n\rangle) + H(\alpha^n) + H(\beta^n).
        \end{multline*}

We will prove the following inequality
        $$
H(\xi) + H(\gamma) - H(\langle\gamma,\alpha^n\rangle)\ge
-c\varepsilon n
        $$
for some constant $c$ that does not depend on $\varepsilon$ and
for sufficiently large $n$. Using this inequality and the symmetric
inequality
        $$
H(\eta) + H(\gamma) - H(\langle\gamma,\beta^n\rangle)
\ge -c\varepsilon n
        $$
we conclude that
        \begin{multline*}
H(\xi)+H(\eta)+(2-2^{-k})H(\gamma) \ge{}\\
{}\ge H(\alpha^n)+ H(\beta^n)-2c\varepsilon n.
        \end{multline*}
Recall that values of $\xi$ are $(v+\delta)n$-bit
strings; therefore $H(\xi)\le (v+\delta) n$.
Using similar arguments for $\eta$ and $\gamma$ and
recalling that $H(\alpha^n)=nH(\alpha)$ and $H(\beta^n)=nH(\beta)$
(independence) we conclude that
        \begin{multline*}
(v+\delta)n+(w+\delta)n+(2-2^{-k})(u+\delta)n\ge{}\\
{}\ge nH(\alpha)+nH(\beta)-2c\varepsilon n.
        \end{multline*}
Dividing over $n$ and recalling that $\varepsilon$ and $\delta$
may be chosen arbitrarily small (according to the definition of
the rate region), we get the statement of Theorem~4.

It remains to prove that
        $$
H(\xi) + H(\gamma) - H(\langle\gamma,\alpha^n\rangle)
\ge -c\varepsilon n
        $$
for some $c$ that does not depend on $\varepsilon$ and for
sufficiently large~$n$. For that we need the following
simple bound:

\begin{lemma}
Let $\mu$ and $\mu'$ be two random variables that coincide with
probability $(1-\varepsilon)$ where $\varepsilon<1/2$. Then
        $$
H(\mu')\le H(\mu)+1+\varepsilon\log m
        $$
where $m$ is the number of possible values of $\mu'$.
\end{lemma}

\smallskip

\begin{proof}
Consider a new random variable $\sigma$ with $m+1$ values that is
equal to $\mu'$ if $\mu\ne\mu'$ and takes a special value if
$\mu=\mu'$. We can use at most $1+\varepsilon\log m$ bits on
average to encode $\sigma$ ($\log m$ bits with probability
$\varepsilon$, if $\mu\ne\mu'$, and one additional bit to
distinguish between the cases $\mu=\mu'$ and $\mu\ne\mu'$).
Therefore, $H(\sigma)\le 1+\varepsilon\log m$. If we know the
values of $\mu$ and $\sigma$, we can determine the value of
$\mu'$, therefore
        $$
H(\mu')\le H(\mu)+H(\sigma)\le H(\mu)+1+\varepsilon\log m.
        $$
\end{proof}

The statement of Lemma~12 remains true if $\mu'$ can be
reconstructed from $\mu$ with probability at least
$(1-\varepsilon)$ (just replace $\mu$ with a function of $\mu$).

Now recall that the pair $\langle\gamma,\alpha^n\rangle$ can be
reconstructed from $\xi$ and $\gamma$ (using the decoding
function $r$) with probability $(1-\varepsilon)$. Therefore,
$H(\langle\gamma,\alpha^n\rangle)$ does not exceed
$H(\langle\xi,\gamma\rangle)+1+c\varepsilon n$ (for some $c$
and large enough~$n$) because both $\gamma$ and $\alpha^n$ have
range of cardinality $O(1)^n$. It remains to note that
$H(\langle\xi,\gamma\rangle)\le H(\xi)+H(\gamma)$.
\end{proof}

\section*{Acknowledgements}
We thank participants of the Kolmogorov seminar, and especially
Alexander Shen and Nikolai Vereshchagin for the formulation of the problem,
helpful discussions and comments.

We wish to thank Emily Cavalcanti, Daniel J. Webre
and the referees for useful comments and suggestions.
\nocite{*}
\bibliographystyle{IEEE}

\end{document}